\documentstyle[prc,aps,epsf,eqsecnum,mbf,preprint]{revtex}
\tightenlines
\def\eq#1{Eq.\ (\ref{#1})}

\def\SM3{\Sigma N (3/2)}
\def\SN1{\Sigma N (1/2)}

\def\TS1{\hbox{}^3S_1}
\def\TD1{\hbox{}^3D_1}

\def\VRGM{V^{\rm RGM}(\varepsilon)}

\begin{document}
\draft

\title   {
Triton binding energy calculated from the $SU_6$ quark-model
nucleon-nucleon interaction
         }

\author  {
Y. Fujiwara$^1$, K. Miyagawa$^2$, M. Kohno$^3$, Y. Suzuki$^4$,
and H. Nemura$^5$
         }

\address {
$^1$Department of Physics, Kyoto University,
Kyoto 606-8502, Japan \\
$\hbox{}^{2}$Department of Applied Physics, Okayama Science University,
Okayama 700-0005, Japan \\
$\hbox{}^{3}$Physics Division, Kyushu Dental College,
Kitakyushu 803-8580, Japan \\
$\hbox{}^{4}$Department of Physics,
Niigata University, Niigata 950-2181, Japan \\
$\hbox{}^{5}$Institute of Particle and Nuclear Physics, KEK,
Tsukuba 305-0801, Japan
}

\maketitle

\bigskip
\bigskip

\begin{abstract}
Properties of the three-nucleon bound state are examined in the Faddeev
formalism, in which the quark-model nucleon-nucleon interaction
is explicitly incorporated to calculate the off-shell $T$-matrix.
The most recent version, fss2, of the Kyoto-Niigata quark-model potential
yields the ground-state energy $E(\hbox{}^3\hbox{H})=-8.514$ MeV
in the 34 channel calculation, when the $np$ interaction is used
for the nucleon-nucleon interaction.
The charge root mean square radii of the $\hbox{}^3\hbox{H}$ and
$\hbox{}^3\hbox{He}$ are 1.72 fm and 1.90 fm, respectively,
including the finite size correction of the nucleons.
These values are the closest to the experiments
among many results obtained by detailed Faddeev calculations
employing modern realistic nucleon-nucleon interaction models.
\end{abstract}

\bigskip

\pacs{21.45.+v, 21.30.-x, 13.75.Cs, 12.39.Jh}

\section{Introduction}

All the present-day quark-model descriptions
of nucleon-nucleon ($NN$) and hyperon-nucleon ($YN$) interactions
incorporate important roles of the quark-gluon degrees of freedom
in the short-range region and the meson-exchange processes
dominating in the medium- and long-range parts
of the interaction. \cite{OY84}
For example, the Kyoto-Niigata quark-model potential employs
a one-gluon exchange Fermi-Breit interaction
and effective meson-exchange potentials (EMEP's) acting between quarks,
and has achieved accurate descriptions of the $NN$ and $YN$ interactions
with a limited number of parameters. \cite{FU96a,FU96b,FU01a,FU01b}
The early version, the model FSS \cite{FU96a,FU96b},
includes only the scalar (S) and pseudoscalar (PS) meson-exchange
potentials as the EMEP's. This model is superseded by the new
model fss2 \cite{FU01b}, which has also introduced
the vector (V) meson-exchange potentials
and the momentum-dependent Bryan-Scott terms included
in the S and V meson EMEP's.
Owing to the introduction of the V mesons,
the model fss2 in the $NN$ sector has attained the accuracy
comparable to that of one-boson exchange potential (OBEP) models.  
For example, the $\chi^2$ values defined by $\chi^2=\sum_{i=1}^N
\left( \delta_i^{cal}-\delta_i^{exp}\right)^2/N$ for
the $J \leq 2$ $np$ phase-shift parameters in the energy
range $T_{\rm lab}=25$ - 300 MeV are $\sqrt{\chi^2}$
= $0.59^\circ$ for fss2, which is compared
with the values \cite{MA89} $1.10^\circ$,
$1.40^\circ$ and $1.32^\circ$ for OBEP, Paris and Bonn
potentials, respectively.
The incorporation of the momentum-dependent Bryan-Scott term
is favorable in extending our quark-model description
of the $NN$ scattering at the non-relativistic
energies to the higher energies up
to $T_{\rm lab}=800$ MeV, and also in describing
reasonable asymptotic behavior of the nucleon s.p. potentials
in the high-momentum region. The agreement of the
higher partial waves up to $J$=4 with the phase shift analysis
is also improved.
In both models FSS and fss2, the existing data
for the $YN$ scattering are well reproduced
and the essential feature of the $\Lambda N$-$\Sigma N$ coupling
remains almost unchanged.
Fixing the model parameters in the strangeness $S=0$ and $-1$
sectors, we proceed to explore interactions for 
any arbitrary combinations of octet baryons ($B_8$). \cite{FU01a}
The $B_8 B_8$ interactions in $S=-2,~-3$ and $-4$ sectors
include the $\Lambda \Lambda$ and $\Xi N$ interactions,
which are recently attracting much interest in the
rapidly developing field of the hypernuclei and
the strangeness nuclear matter. 
The interaction derived in these models
may be used for realistic calculations
in few-baryon systems, like the triton $\hbox{}^3\hbox{H}$,
the hypertriton $\hbox{}^3_\Lambda \hbox{H}$,
and also in various types of baryonic matter.
This project, however, involves a non-trivial problem
of determining how to extract the effective two-baryon interaction
from the microscopic quark-exchange kernel.
The basic baryon-baryon interaction is formulated as a composite-particle
interaction in the framework of the resonating-group method (RGM).
If we rewrite the RGM equation in the form
of a Schr{\" o}dinger-type equation,
the interaction term becomes non-local and energy dependent.
Furthermore, the RGM equation sometimes involves
redundant components, due to the effect of
the antisymmetrization, which is related to the existence
of the Pauli-forbidden states. 
In such a case, the full off-shell $T$-matrix is not well
defined in the standard procedure, which usually assumes
simple energy-independent local potentials. \cite{GRGM}
Since these features are related to the description
of the short-range part in the quark model,
it would be desirable if the quark-exchange kernel could be used directly
in application to many-baryon systems.

In this paper, we will show some results of the Faddeev
calculation which directly employs the quark-model $NN$ interactions
fss2 and FSS to derive the off-shell $T$-matrix.
Following the notation in ref.\cite{GRGM,TRGM},
we write the RGM equation of the $(3q)$-$(3q)$ system
in the form of the Schr{\" o}dinger-type equation
\begin{eqnarray}
\left[\,\varepsilon-h_0-V^{\rm RGM}(\varepsilon)\,\right] \chi=0\ ,
\label{form1}
\end{eqnarray}
where $\varepsilon$ is the total energy in the center-of-mass system,
measured from the two-cluster threshold, $\varepsilon=E-2E_N$,
$h_0$ is the kinetic-energy operator of the $NN$ relative motion, and
\begin{eqnarray}
\VRGM=V_{\rm D}+G+\varepsilon K\ ,
\label{form2}
\end{eqnarray}
is the RGM kernel composed of the direct potential $V_{\rm D}$, the sum
of the exchange kinetic-energy and interaction kernels,
$G=G^{\rm K}+G^{\rm V}$,
and the exchange normalization kernel $K$.
Since there is no Pauli forbidden state in the $NN$ system,
we can solve the Lippmann-Schwinger equation
\begin{eqnarray}
T(\omega,\varepsilon)=\VRGM+\VRGM g_0(\omega)
T(\omega,\varepsilon)\ ,
\label{form3}
\end{eqnarray}
with $g_0(\omega)=1/(\omega-h_0+i0)$, by assuming
the $2N$ energy $\varepsilon$ as a mere parameter.
The Faddeev equation for the $3N$ bound state is
given by the eigenvalue problem
\begin{eqnarray}
\lambda(E) \psi_\alpha
=G^{(+)}_0(E) T^{(3)}_\alpha (E, \varepsilon_\alpha)
(\psi_\beta+\psi_\gamma)\ ,
\label{form4}
\end{eqnarray}
with $\lambda(E)=1$, where the two-body $T$-matrix
in the three-body model space is given by
\begin{eqnarray}
T^{(3)}_\alpha(E, \varepsilon_\alpha)
=T_\alpha(E-h_{0\bar{\alpha}}, \varepsilon_\alpha)\ ,
\label{form5}
\end{eqnarray}
and $G_0(E)=1/(E-H_0+i0)$ is the free Green's function for the three-body
kinetic-energy operator $H_0=h_{0\alpha}+h_{0\bar{\alpha}}$.
The energy dependence of the two-cluster RGM kernel is
self-consistently determined \cite{TRGM,RED} through
\begin{eqnarray}
\varepsilon_\alpha=\langle \Psi |
h_{0\alpha}+V^{\rm RGM}_\alpha(\varepsilon_\alpha) | \Psi \rangle
={1 \over 3}E+{1 \over 2} \langle \varphi_\alpha |
H_0 |\Psi \rangle\ ,
\label{form6}
\end{eqnarray}
where $\Psi=\varphi_\alpha+\varphi_\beta+\varphi_\gamma$ is
the normalized total wave function for the $3N$ bound state.
In practice, we start from some specific values
of $\varepsilon_\alpha$ and $E$,
and solve \eq{form4} to find a negative three-body energy $E$ such that
the eigenvalue $\lambda(E)$ becomes 1.
The normalized Faddeev component $\varphi_\alpha$ yields
a new value of $\varepsilon_\alpha$ through \eq{form6}.\footnote{As a
system of identical three particles, $\varepsilon_\alpha$,
$\varepsilon_\beta$ and $\varepsilon_\gamma$ are all equal
and are expressed as $\varepsilon$ in Table\,\protect\ref{table3}.} 
Since it is usually not equal to the starting value,
we repeat the process by using the new value.
This process of double iteration converges very fast if
the starting values of $\varepsilon_\alpha$ and $E$ are
appropriately chosen.

For the numerical calculation, we discretize the continuous momentum
variables $p$ and $q$ for the Jacobi coordinate vectors,
using the Gauss-Legendre $n_1$- and $n_2$-point
quadrature formulas, respectively,
for each of the three intervals of 0 - 1 $\hbox{fm}^{-1}$,
1 - 3 $\hbox{fm}^{-1}$ and 3 - 6 $\hbox{fm}^{-1}$. 
The small contribution from the intermediate integral
over $p$ beyond $p_0=6~\hbox{fm}^{-1}$ in the $2N$ $T$-matrix
calculation is also taken into account by using
the Gauss-Legendre $n_3$-point quadrature formula through the
mapping $p=p_0+{\rm tan}(\pi/4)(1+x)$.\footnote{These $n_3$ points
for $p$ are not included for solving the Faddeev
equation (\protect\ref{form4}),
since it causes a numerical inaccuracy for the interpolation.}
The momentum region $q=$ 6 $\hbox{fm}^{-1}$ - $\infty$ is
also discretized by the $n_3$ point formula just
as in the $p$ discretization case. 
We take $n_1$-$n_2$-$n_3$=10-10-5, for which well converged
results are obtained at least for 2 and 5 channel calculations.
The partial-wave decomposition of
the $2N$ RGM kernel is carried out numerically using
the Gauss-Legendre 20-point quadrature formula.
The modified spline interpolation technique
developed in \cite{GL82a} is employed for
constructing the rearrangement matrix.
For the diagonalization of the large non-symmetric matrix,
the Arnordi-Lanczos algorithm recently developed in the ARPACK
subroutine package \cite{AR96} is very useful.

Tables\,\ref{table1} and \ref{table2} list the deuteron properties
and the $NN$ effective range parameters predicted by fss2 and FSS,
respectively.
All the calculations in the present paper are carried out
in the isospin basis.
For a realistic calculation of the $\hbox{}^3\hbox{H}$ binding
energy, it is essential to use the $NN$ interaction that
reproduces the correct $D$-state probability ($P_D$) of the deuteron
and the effective range parameters
of the $\hbox{}^1S_0$ scattering. \cite{Bra88b}
Since all the realistic $NN$ interactions reproduce the $NN$ phase
shifts more or less correctly, the strength of the central attraction is
counterbalanced with that of the tensor force.
Namely, if the interaction has a weaker tensor force, then
it should have a stronger central attraction.
Generally speaking, the effect of the tensor force is
reduced in the nuclear many-body systems, in comparison with
the bare two-nucleon collision.
This implies that the $NN$ interaction with a weaker tensor force
is favorable, in order to obtain sufficient binding energies
of the nuclear many-body systems.
The weak tensor force, however, causes various problems like
a too small value for the deuteron quadrupole moment $Q_d$ and
some disagreement of the mixing parameter $\varepsilon_1$ of
the $\hbox{}^3S_1+\hbox{}^3D_1$ coupling. For example,
the Reid soft core potential (RSC) \cite{Rei68} gives $P_D=6.5~\%$ and
predicts too small $\hbox{}^3\hbox{H}$ binding energy,
$B_t=7.35~\hbox{MeV}$, compared with the experimental
value $B^{\rm exp}_t=8.48~\hbox{MeV}$.
A series of the Bonn potentials reproduce
the $NN$ phase shifts very accurately, but they have a tendency
that the tensor force is generally rather weak. \cite{MA89}
The model C has the strongest tensor force $P_D=5.61~\%$,
yielding $B_t=7.99~\hbox{MeV}$.
The value $P_D$ becomes smaller for models B and A,
and the value of $B_t$ becomes larger, correspondingly.
The following results are given in ref.\,\cite{MA89}:
model-B ($P_D=5.0~\%,~B_t=8.13~\hbox{MeV}$),
model-A ($P_D=4.4~\%,~B_t=8.32~\hbox{MeV}$).
These results are all obtained in the 34 channel calculations
(including the $2N$ total angular momentum $J\leq 4$), 
and by using the $np$ interaction.
In fact, the effects of the charge dependence and
the charge asymmetry are important for the detailed discussion,
and it is estimated to be
about 190 keV in refs.\,\cite{MA89} and \cite{Bra88a}.
The most recent Faddeev calculation employing the CD-Bonn
potential \cite{MA01} incorporates these effect,
and predicts $B_t=8.014~\hbox{MeV}$ \cite{NO97} for $P_D=4.85~\%$.
The present status of the $\hbox{}^3\hbox{H}$ binding energy calculation
is summarized as that more than 0.5 MeV is missing if the
two-nucleon force of any realistic $NN$ interactions
is only employed. \cite{NO00}

On the other hand, our result of $P_D$ in Table\,\ref{table1}
is about 5.5 $\%$ both in the fss2 and FSS cases.
We think that this is a reasonable value, in spite of the
fact that $Q_d$ of fss2 is too small.
This is because a careful evaluation of the meson-exchange current
contributions to $Q_d$, which could be as large
as 0.01 $\hbox{fm}^2$ \cite{AL78,KO83}, must be made.
Our results of the effective range parameters in Tables\,\ref{table2}
are not as perfect as those of the Bonn B potential.
It should be noted that the effects of the higher-order terms
of the Coulomb interaction are not incorporated in these calculations.
The deuteron binding energy and the scattering length $a_s$ for
the $\hbox{}^1S_0$ state are fit in determining our quark-model parameters.

Table\,\ref{table3} list the results of the Faddeev calculations
by fss2 and FSS in various types of truncations of the model space.
The 5 channel calculation with $J \leq 1^+$ incorporates only
the partial waves $\hbox{}^3S_1+\hbox{}^3D_1$ and $\hbox{}^1S_0$ for
the $2N$ $T$-matrix. Similarly, the 18 and 34 channel calculations
incorporate the partial waves with $J \leq 2$ and $J \leq 4$,
respectively.
We find that the energy gain in the 5 channel to 34 channel
calculation is about 330 - 360 keV,
which is the same tendency for the realistic $NN$ potentials
with a strong tensor force, such as the RSC
and Paris potentials \cite{Bra88b}.
The convergence is not enough even in the 34 channel calculation,
and we expect the further energy gain of the order of 10 - 20 keV.
The model fss2 predicts $B_t=8.51~\hbox{MeV}$ and seems to give
too large binding energy, compared with experiment.
In fact, it underbinds by 150 - 160 keV, if the effects
of the charge dependence and the charge asymmetry of the $NN$ interaction
is taken into account.
The scenario assuming the most favorable Bonn A potential
is given in Table 11.1 of ref.\,\cite{MA89}, which tells us that
the corrected value due to the charge dependence
and the charge asymmetry of the two-body
force is 8.13 MeV and the rest, 350 keV, is attributed to
the combined contribution of the three-body force and
the medium effect of the two-body force. 
Our result using the quark-model potentials indicates
that one can reduce the net effect besides the two-nucleon force
to less than half of the OBEP values,
by keeping the deuteron $D$-state probability in a reasonable magnitude.

Note that the $2N$ energy $\varepsilon_\alpha$ in \eq{form6}
is directly related to the separation of the total
energy  $E(\hbox{}^3\hbox{H})$ into the kinetic-energy contribution,
$\langle H_0 \rangle=2 (3 \varepsilon_\alpha-E)$, and the
potential-energy contribution, $\langle V \rangle=3 (E-2 \varepsilon_\alpha)$.
In the 34 channel calculations, these are given by
$\langle H_0 \rangle=43.95$ MeV, $\langle V \rangle=-52.47$ MeV
for fss2, and $\langle H_0 \rangle=41.83$ MeV,
$\langle V \rangle=-50.22$ MeV for FSS.
If we compare these with the results \cite{NO00}
of the CD-Bonn potential ($\langle H_0 \rangle=37.42$ MeV, 
$\langle V \rangle=-45.43$ MeV) and the AV-18 potential
($\langle H_0 \rangle=46.73$ MeV, $\langle V \rangle=-54.35$ MeV),
we find that our quark-model potentials give a moderate amount of
the kinetic-energy contribution just between the CD-Bonn and AV18 potentials. 

Table\,\ref{table3} also shows the calculated
charge root mean square (rms) radii of $\hbox{}^3{\rm H}$ and
$\hbox{}^3{\rm He}$ obtained by fss2 and FSS. The finite size corrections
of the nucleons are made through
\begin{eqnarray}
& & \langle r^2\rangle_{\hbox{}^3{\rm H}}={R_C(\hbox{}^3{\rm H})}^2
+(0.8502)^2-2\times (0.3563)^2\ ,\nonumber \\
& & \langle r^2 \rangle_{\hbox{}^3{\rm He}}={R_C(\hbox{}^3{\rm He})}^2
+(0.8502)^2-{1 \over 2}\times (0.3563)^2\ ,
\label{form7}
\end{eqnarray}
where ${R_C}^2$ stands for the square of the charge rms radius
for the point nucleons. Since our $3N$ bound state wave functions
are given in the momentum representation, we first calculate
the charge form factors $F_C(Q^2)$, according to the formulation given
in ref.\,\cite{GL82b}. ${R_C}^2$ is then extracted from
the power series expansion of $F_C(Q^2)$ with respect to $Q^2$.
We have employed 20 points, $Q=0.05 \times n~\hbox{fm}^{-1}$
with $n=1~\hbox{-}~20$, for the extrapolation to $Q=0$.
In the present calculation, the Coulomb force and the relativistic
correction terms \cite{KI88} of the charge current operator
are entirely neglected.
The experimental values are rather difficult to
determine, as discussed in ref.\,\cite{KI88}. Here we compare
our results with two empirical values
\begin{eqnarray}
\sqrt{\langle r^2\rangle_{\hbox{}^3{\rm H}}}=\left\{ \begin{array}{c}
1.70\pm 0.05~\hbox{fm} \quad \protect\cite{CO65} \\
1.81\pm 0.05~\hbox{fm} \quad \protect\cite{MA86} \\
\end{array}\right.\ \ ,\qquad
\sqrt{\langle r^2\rangle_{\hbox{}^3{\rm He}}}=\left\{ \begin{array}{c}
1.87\pm 0.05~\hbox{fm} \quad \protect\cite{CO65} \\
1.93\pm 0.03~\hbox{fm} \quad \protect\cite{MA86} \\
\end{array}\right.\ \ .
\label{form8}
\end{eqnarray}
We find that the agreement with the experiment is satisfactory
both for fss2 and FSS.

The Faddeev calculations for $\hbox{}^3\hbox{H}$, 
using the quark-model $NN$ potentials,
have been carried out by Takeuchi, Cheon and Redish \cite{TA92},
and recently by the Salamanca-J{\" u}lich group \cite{SA01}.
In the former calculation, the model QCM-A, by the Tokyo University
group gives the $NN$ phase shifts with almost the same accuracy
as our model FSS.
The model QCM-A predicts $P_D=5.58~\%$ for the deuteron $D$-state
probability and $B_t=8.01~\hbox{-}~ 8.02~\hbox{MeV}$ for
the $\hbox{}^3\hbox{H}$ binding energy in the 5 channel
calculation. This is very similar to our results for the model FSS. 
On the other hand, the Salamanca-J{\" u}lich group
predicts $B_t=7.72~\hbox{MeV}$, in spite of the very small $D$-state
probability $P_D=4.85~\%$. It is not clear to us how they
treated the energy dependence of the RGM kernel at the process
of the separable expansion for solving the Faddeev
equation. They have to improve the fit of the $NN$ phase shifts
for higher partial waves (especially, for the $P$ waves),
in order to extend their calculation to more than 5 channels. 

In summary, we have carried out the Faddeev calculation
for the three-nucleon bound state, by explicitly incorporating
the off-shell $T$-matrix derived from the RGM exchange kernel
of the quark-model $NN$ interaction.
The energy dependence of the two-cluster RGM kernel is
self-consistently treated. \cite{TRGM,RED}
For the two models fss2 \cite{FU96a,FU96b} and FSS \cite{FU01a,FU01b}, 
we have obtained $E(\hbox{}^3\hbox{H})=-8.514$ MeV (fss2), and
$-8.390$ MeV (FSS) in the 34 channel calculation
using the $np$ interaction.
The charge rms radii of the $\hbox{}^3\hbox{H}$ and $\hbox{}^3\hbox{He}$
are in fair agreement with experiment:
$\sqrt{\langle r^2\rangle_{\hbox{}^3{\rm H}}}=1.72$ fm (fss2),
1.74 fm (FSS) and $\sqrt{\langle r^2\rangle_{\hbox{}^3{\rm He}}}
=1.90$ fm (fss2), 1.92 fm (FSS).
In these calculations, the Coulomb force and the relativistic correction
terms are neglected.
In view of the fact that the $NN$ phase shifts of FSS are
not that excellent, the results of fss2 are more meaningful.
These numbers are the closest to the experiments
among many results obtained by Faddeev calculations
employing modern realistic $NN$ interaction models.
Since both models fss2 and FSS have a common feature
in describing the short range correlation by the quark exchange
kernel, it is important to clarify the mechanism in which
the quark-model potentials give larger $\hbox{}^3\hbox{H}$ binding
energy than the meson-exchange potentials.
The off-shell behavior of the RGM $T$-matrix is closely connected
to this alternative description of the short-range correlations.
More detailed study on this point is now under way.


\acknowledgments

This research is supported by Japan Grant-in-Aid for Scientific
Research from the Ministry of Education, Science, Sports and
Culture (12640265, 14540249).

\nopagebreak


\begin{table}[h]
\caption{The deuteron properties by fss2 and FSS in the isospin basis.
The results by the Bonn B potential \protect\cite{MA89} are
also shown for comparison. A small difference in FSS from Table IV
of \protect\cite{FU96b} is due to the numerical inaccuracy
in the previous calculation.
The effect of the meson exchange current is not included
in the calculated values of $Q_d$ and $\mu_d$.
}
\label{table1}
\begin{center}
\renewcommand{\arraystretch}{1.2}
\setlength{\tabcolsep}{3mm}
\begin{tabular}{cccccc}
 & FSS & fss2 & Bonn B & Expt. & Ref. \\
\hline
$\epsilon_d$ (MeV) & 2.256  & 2.225  & 2.2246 & $2.224644 \pm 0.000046$
 & \protect\cite{DU83} \\
$P_D$ ($\%$)       & 5.86   & 5.49   & 4.99   & $-$ & \\
$\eta=A_D/A_S$     & 0.0267 & 0.0253 & 0.0264 & $0.0256 \pm 0.0004$
 & \protect\cite{RO90} \\
rms (fm) & 1.963   & 1.960  & 1.968  & $1.9635 \pm 0.0046$
 & \protect\cite{DU83} \\
$Q_d$ (fm$\hbox{}^2$) & 0.283 & 0.270 & 0.278 & $0.2860 \pm 0.0015$
 & \protect\cite{BI79} \\
$\mu_d$ ($\mu_N$) & 0.8464  & 0.8485 & 0.8514 &
$0.857406 \pm 0.000001$ & \protect\cite{Lin65} \\
\end{tabular}
\end{center}
\end{table}


\begin{table}[h]
\caption{The $NN$ effective range parameters calculated
by fss2 and FSS in the isospin basis.
The results by the Bonn B potential \protect\cite{MA89} are
also shown for comparison.
The higher-order terms of the Coulomb force are not included.
The experimental values are taken from \protect\cite{DU83}.
}
\label{table2}
\begin{center}
\renewcommand{\arraystretch}{1.2}
\setlength{\tabcolsep}{3mm}
\begin{tabular}{ccccc}
 & FSS & fss2 & Bonn B & Expt. \\
\hline
$a_s$ (fm) & $-23.64$ & $-23.76$ & $-23.75$ & $-23.748 \pm 0.010$ \\
$r_s$ (fm) &   2.62   & 2.58     & 2.71     & $2.75 \pm 0.05$     \\
$a_s$ (fm) &   5.41   & 5.399    & 5.424    & $5.424 \pm 0.004$   \\
$r_s$ (fm) &   1.76   & 1.730    & 1.761    & $1.759 \pm 0.005$   \\
\end{tabular}
\end{center}
\end{table}

\begin{table}[h]
\caption{
The three-nucleon bound state properties predicted by the Faddeev
calculation with fss2 and FSS. The $np$ interaction is used
in the isospin basis. 
The discretization points of $p$ and $q$ are specified
by the values of $n_1$-$n_2$-$n_3$=10-10-5 (see the text).
The column ``channels'' implies the number of two-nucleon
channels included, and $n_{\rm max}=n(3n_1)(3n_2+n_3)$ for the
$n$ channel calculation is the dimension
of the diagonalization for the Faddeev equation.
$E(\hbox{}^3\hbox{H})$ is the ground state
energy, and $\protect\sqrt{\langle r^2\rangle_{\hbox{}^3{\rm H}}}$ and
$\protect\sqrt{\langle r^2\rangle_{\hbox{}^3{\rm He}}}$ are the charge
rms radii for $\hbox{}^3{\rm H}$ and $\hbox{}^3{\rm He}$,
respectively, with the proton and neutron size corrections
introduced by \protect\eq{form7}.
The Coulomb force and the relativistic corrections are neglected.
$\varepsilon$ is the $2N$ expectation value,
\protect\eq{form6}, determined self-consistently.
}
\label{table3}
\begin{center}
\renewcommand{\arraystretch}{1.2}
\setlength{\tabcolsep}{2mm}
\begin{tabular}{ccccccc}
model & channels & $n_{\rm max}$ & $\varepsilon(2N)$
 & $E(\hbox{}^3\hbox{H})$
 & $\protect\sqrt{\langle r^2\rangle_{\hbox{}^3{\rm H}}}$
 & $\protect\sqrt{\langle r^2\rangle_{\hbox{}^3{\rm He}}}$ \\
 & & & (MeV) & (MeV)  & (fm) & (fm) \\
\hline
     &  2 ch &  2,100 & 2.361 & $-7.807$ & 1.80 & 1.96 \\
     &  5 ch &  5,250 & 4.341 & $-8.189$ & 1.75 & 1.92 \\
fss2 & 10 ch & 10,500 & 4.249 & $-8.017$ & 1.76 & 1.94 \\
     & 18 ch & 18,900 & 4.460 & $-8.439$ & 1.72 & 1.90 \\
     & 34 ch & 35,700 & 4.488 & $-8.514$ & 1.72 & 1.90 \\
\hline
     &  2 ch &  2,100 & 2.038 & $-7.674$ & 1.83 & 1.99 \\
     &  5 ch &  5,250 & 3.999 & $-8.034$ & 1.78 & 1.95 \\
FSS  & 10 ch & 10,500 & 3.934 & $-7.909$ & 1.78 & 1.97 \\
     & 18 ch & 18,900 & 4.160 & $-8.342$ & 1.74 & 1.93 \\
     & 34 ch & 35,700 & 4.175 & $-8.390$ & 1.74 & 1.92 \\
\end{tabular}
\end{center}
\end{table}

\end{document}